\documentstyle[multicol,eqsecnum,aps,psfig]{revtex}
\renewcommand{\narrowtext}{\begin{multicols}{2}
\global\columnwidth20.5pc\noindent}
\renewcommand{\widetext}{\end{multicols}
\global\columnwidth42.5pc}
\begin{document}
\draft
\preprint{29 April 2003}
\title{Nuclear spin-lattice relaxation in ferrimagnetic
       clusters and chains:\\
       A contrast between zero and one dimensions}
\author{Hiromitsu Hori and Shoji Yamamoto}
\address{Division of Physics, Hokkaido University,
         Sapporo 060-0810, Japan}
\date{29 April 2003}
\maketitle
\begin{abstract}
Motivated by ferrimagnetic oligonuclear and chain compounds synthesized by
Caneschi {\it et al.}, both of which consist of alternating manganese(II)
ions and nitronyl-nitroxide radicals, we calculate the nuclear
spin-lattice relaxation rate $1/T_1$ employing a recently developed
modified spin-wave theory.
$1/T_1$ as a function of temperature drastically varies with the location
of probe nuclei in both clusters and chains, though the relaxation time
scale is much larger in zero dimension than in one dimension.
$1/T_1$ as a function of an applied field in long chains forms a striking
contrast to that in finite clusters, diverging with decreasing field like
inverse square root at low temperatures and logarithmically at high
temperatures.
\end{abstract}
\narrowtext

\section{Introduction}

   Clusters of metal ions \cite{G1054} serve to test the validity of
quantum mechanical approaches at the nanometer scale.
Among others a dodecanuclear manganese complex of formula
[Mn$_{12}$O$_{12}$(CH$_3$COO)$_{16}$(H$_2$O)$_4$] \cite{L2042} and
an octanuclear iron complex of formula
[Fe$_8$(N$_3$C$_6$H$_{15}$)$_6$O$_2$(OH)$_{12}$]$^{8+}$ \cite{W77},
which are hereafter abbreviated as Mn$_{12}$ and Fe$_8$, respectively,
have been attracting considerable interest in this context.
In these clusters, magnetic hysteresis curves are independent of
temperature and consist of equally separated steps at sufficiently low
temperatures \cite{F3830,T145,S4645,W133}, suggesting quantum tunneling
of the magnetization.
Both clusters have a magnetic ground state of total spin $10$ and are
often treated as a rigid spin-$10$ object with an Ising-type
magneto-crystalline anisotropy.
Such a phenomenological interpretation is indeed successful for the
magnetic relaxation at sufficiently low temperatures and weak fields, but
it masks individual internal structures of various magnetic clusters.
Without a microscopic Hamiltonian, we could not essentially distinguish
the Mn$_{12}$ and Fe$_8$ clusters.
We cannot compare nanomagnets with bulk magnets in the same microscopic
language until we describe them in terms of their constituent ion spins.

\begin{figure}
\centerline
{\mbox{\psfig{figure=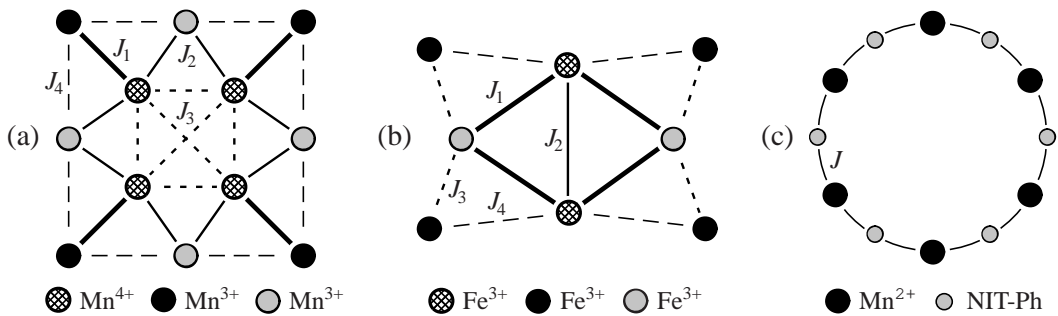,width=140mm,angle=0}}}
\vspace*{2mm}
\caption{Schematic plot of Mn$_{12}$ (a), Fe$_8$ (b), and
         (MnNIT)$_6$ (c).
         Symmetry-inequivalent sites are distinguishably drawn.}
\label{F:illust}
\end{figure}

   Thus motivated, several authors made an attempt to estimate exchange
interactions in the Mn$_{12}$ and Fe$_8$ clusters calculating the
low-lying energy spectra \cite{S1804,R064419,B184435}, magnetic
susceptibilities \cite{D2264}, magnetization curves \cite{Z1140,R054409},
inelastic neutron-scattering spectra \cite{K6919}, tunneling splittings
\cite{D392}, and nuclear spin-lattice relaxation rates \cite{Y157603}.
The $8$-spin modeling of the Mn$_{12}$ cluster \cite{K6919,D392} revealed
the decisive role of multispin effects in magnetic tunneling, while the
extended {\it ab initio} calculation \cite{B184435} demonstrated the
relevance of metal-ligand orbital bybridization to intramolecular exchange
interactions.
However, the predicted magnetic structures are rather controversial for
both clusters.
The Mn$_{12}$ cluster consists of eight Mn$^{3+}$ ions of spin $2$ and
four Mn$^{4+}$ ions of spin $\frac{3}{2}$, while the Fe$_8$ cluster
contains eight Fe$^{3+}$ ions, both of which are coupled to each other
through four kinds of exchange interactions (Fig. \ref{F:illust}).
Antisymmetric exchange interactions of the Dzyaloshinsky-Moriya type are
also assumed to be relevant to the Mn$_{12}$ cluster \cite{K6919}.
There is no established way of assigning the magnetic anisotropy to each
ion site.
Thus, complicated intracluster magnetic structures block our microscopic
understanding of mesoscopic magnetism.
Nanomagnets such as Mn$_{12}$ and Fe$_8$ have indeed been providing
fascinating observations but are not necessarily suitable for a
comparative study on zero- and one (or higher)-dimensional quantum
magnetism.

\widetext
\vskip 2mm
\narrowtext

   In order to reveal how paramagnetic spins grow into bulk magnets, we
take a great interest in ferrimagnetic ring clusters \cite{C2795} and
chains \cite{C1756} synthesized by Caneschi {\it et al.}, both of which
consist of manganese hexafluoroacetylacetonates (hfac) and nitronyl
nitroxide radicals
2-$R$-4,4,5,5-tetramethyl-4,5-dihydro-1$H$-imidazolyl-1-oxyl 3-oxide
(NIT-$R$) with
$R=\mbox{phenyl (Ph)}$, $\mbox{isopropyl ($i$-Pr)}$,
   $\mbox{ethyl (Et)}$, $\mbox{methyl (Me)}$.
The [Mn(hfac)NIT-Ph]$_6$ cluster, hereafter abbreviated as (MnNIT)$_6$,
has antiferromagnetic exchange coupling between the Mn$^{2+}$ ions of spin
$\frac{5}{2}$ and the radicals of spin $\frac{1}{2}$ so as to exhibit a
magnetic ground state of total spin $12$ (Fig. \ref{F:illust}).
The [Mn(hfac)NIT-$R$]$_\infty$ chain, hereafter abbreviated as
(MnNIT)$_\infty$, may be regarded as a one-dimensional analog of
(MnNIT)$_6$.
Their simple magnetic structures, describable within isotropic exchange
Hamiltonians \cite{C2795,C1756}, are suitable enough to compare
oligonuclear ferrimagnets with those of one dimension in their intrinsic
features.

   The theoretical tool we employ here is a recently developed modified
spin-wave theory, which is quite useful in understanding thermal
\cite{Y14008,O8067,N214418} as well as ground-state \cite{B3921,N1380}
properties of various one-dimensional ferrimagnets.
We inquire further into zero dimension and dynamic properties.
Besides direct observations of resonant magnetization tunneling,
nuclear-magnetic-resonance (NMR) measurements have extensively been
performed for cluster magnets.
$^1$H, $^2$D, $^{13}$C, and $^{55}$Mn NMR in Mn$_{12}$
\cite{L514,A2941,F14246,A064420,K224425}, $^1$H and $^2$D NMR in Fe$_8$
\cite{F094439,U073309}, and $^1$H NMR in a hexanuclear copper complex
\cite{F6265} significantly contributed toward revealing the quantum
dynamics of cluster magnets.
As for the systems of Mn(hfac)NIT-$R$, we may consider NMR measurements
using as probes $^1$H, $^{13}$C, $^{19}$F, and $^{55}$Mn nuclei.
We calculate the nuclear spin-lattice relaxation rate 1/$T_1$ as a
function of temperature, an applied field, and the location of probe nuclei.

\section{Model Hamiltonian}

\begin{figure}
\centerline
{\mbox{\psfig{figure=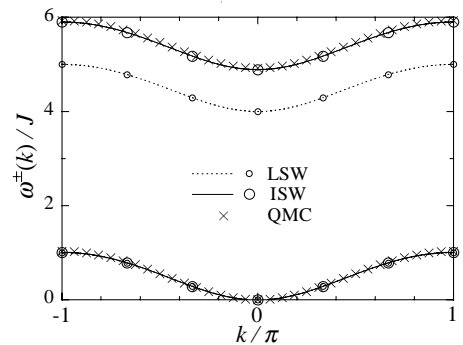,width=60mm,angle=0}}}
\vskip 2mm
\caption{Dispersion relations of the linear (LSW) and interacting (ISW)
         spin waves,
         $\omega_1^\pm(k)$ and $\omega_1^\pm(k)+\omega_0^\pm(k)$, at
         $N=6$ and $N=\infty$ under zero field.
         Corresponding quantum Monte Carlo calculations (QMC) at $N=32$
         are also shown for reference.}
\label{F:dsp}
\end{figure}

   (MnNIT)$_6$ and (MnNIT)$_\infty$ are both described by an isotropic
spin-$(S,s)$ Heisenberg Hamiltonian \cite{C2795,C1756}
\begin{equation}
   {\cal H}
      =J\sum_{i=1}^N
        \left(
         \mbox{\boldmath$S$}_{i} \cdot \mbox{\boldmath$s$}_{i}
        +\mbox{\boldmath$s$}_{i} \cdot \mbox{\boldmath$S$}_{i+1}
        \right)
      -g\mu_{\rm B} H\sum_{i=1}^N(S_i^z+s_i^z)\,,
   \label{E:H}
\end{equation}
where $S=\frac{5}{2}$, $s=\frac{1}{2}$, and we have set their $g$ factors
both equal to $g$.
Introducing bosonic operators for the spin deviation in each sublattice
via
$S_i^+=(2S-a_i^\dagger a_i)^{1/2}a_i$,
$S_i^z=S-a_i^\dagger a_i$,
$s_i^+=b_i^\dagger(2s-b_i^\dagger b_i)^{1/2}$,
$s_i^z=-s+b_i^\dagger b_i$,
we expand the Hamiltonian with respect to $1/S$ as
\begin{equation}
   {\cal H}=-2SsJN+{\cal H}_1+{\cal H}_0+O(S^{-1})\,,
   \label{E:Hexp}
\end{equation}
where we assume that $O(S)=O(s)$ and ${\cal H}_i$ is the $O(S^i)$
contribution.
Considering the perturbational treatment of ${\cal H}_0$ to ${\cal H}_1$,
we obtain the diagonal spin-wave Hamiltonian as
\begin{equation}
   {\cal H}
   =E_{\rm g}
   +\sum_k
    \left[
     \omega^-(k)\alpha_k^\dagger\alpha_k
    +\omega^+(k)\beta_k^\dagger\beta_k
    \right]
   +O(S^{-1})\,,
\end{equation}
with $E_{\rm g}=-2SsJN+\sum_{i=1,0}E_i$ and
$\omega^\pm(k)=\sum_{i=1,0}\omega_i^\pm(k)$,
where $E_i$ and $\omega_i^\pm(k)$ are the $O(S^i)$ corrections to the
ground-state energy and the dispersion relations, respectively.
The dispersions of the linear spin waves, $\omega_1^\pm(k)$, and the
corrections due to the interactions between them, $\omega_0^\pm(k)$, are,
respectively, given by
\begin{eqnarray}
   &&
   \omega_1^\pm(k)=[\omega_k\pm(S-s)]J\mp g\mu_{\rm B}H\,,
   \label{E:dsp1}\\
   &&
   \omega_0^\pm(k)
   =-2(S+s)J{\mit\Gamma}_1\frac{\sin^2(k/2)}{\omega_k}
   \nonumber\\
   &&\qquad\quad\ 
   +\frac{J{\mit\Gamma}_2}{\sqrt{Ss}}[\omega_k\pm(S-s)]\,,
   \label{E:dsp0}
\end{eqnarray}
where
$\omega_k=[(S-s)^2+4Ss\sin^2(k/2)]^{1/2}$,
${\mit\Gamma}_1=-(1/2N)\sum_k[1-(S+s)/\omega_k]$, and
${\mit\Gamma}_2=(1/N)\sum_k(\sqrt{Ss}/\omega_k)\cos^2(k/2)$.
${\mit\Gamma}_1$ is nothing but the quantum spin reduction
$(1/N)\sum_i\langle a_i^\dagger a_i\rangle_{T=0}
=(1/N)\sum_i\langle b_i^\dagger b_i\rangle_{T=0}$
and can analytically be evaluated as $1/\sqrt{31}+1/\sqrt{21}-7/24$ at
$N=6$, which is slightly larger than the $N\rightarrow\infty$ numerical
estimate $0.106139$ and thus suggests growing quantum fluctuations with
decreasing system size.
In Fig. \ref{F:dsp} we plot the spin-wave dispersions together with
quantum Monte Carlo calculations.
The lower branch, which reduces the ground-state magnetization, is of
ferromagnetic aspect exhibiting a quadratic dispersion at small momenta,
whereas the upper branch, which enhances the ground-state magnetization,
is of antiferromagnetic aspect being gapped from the ground state.
The antiferromagnetic mode is remarkable for its $O(S^0)$ quantum
correction.
\vspace*{-35mm}

\begin{figure}
\centerline
{\mbox{\psfig{figure=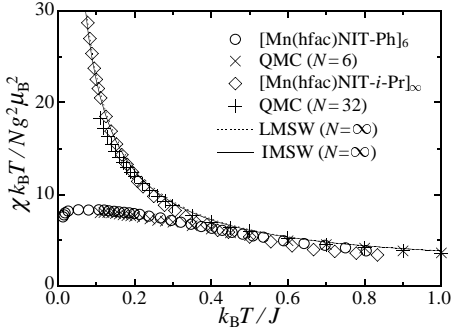,width=60mm,angle=0}}}
\vskip 2mm
\caption{Magnetic-susceptibility measurements on [Mn(hfac)NIT-Ph]$_6$ at
         $H=0.5\,\mbox{T}$ [15] and on [Mn(hfac)NIT-$i$-Pr]$_\infty$ at
         $H=0$ [16] compared with quantum Monte Carlo calculations
         (QMC).
         They are in good agreement assuming
         $J/k_{\rm B}\simeq 370\,\mbox{K}$ for [Mn(hfac)NIT-Ph]$_6$ and
         $J/k_{\rm B}\simeq 360\,\mbox{K}$ for
         [Mn(hfac)NIT-$i$-Pr]$_\infty$.
         Linear (LMSW) and interacting (IMSW) modified spin-wave
         calculations at $N=\infty$ are also shown.}
\label{F:chiT}
\end{figure}

\section{Modified Spin-Wave Formalism}

   In order to calculate thermodynamics, we consider modifying the
spin-wave scheme \cite{T168}, that is, controlling the number of bosons
induced thermally.
For isotropic ferrimagnets, we claim that the thermal fluctuation should
cancel the staggered magnetization as \cite{Y14008}
\begin{equation}
   (S+s)\sum_k\sum_{\sigma=\pm}\frac{\bar{n}_k^\sigma}{\omega_k}
  =(S+s)N\,,
   \label{E:const}
\end{equation}
where
$\bar{n}_k^\sigma=\sum_{n^\pm=0}^\infty n^\sigma P_k(n^-,n^+)$ with
$P_k(n^-,n^+)$ being the probability of $n^-$ ferromagnetic and $n^+$
antiferromagnetic spin waves appearing in the $k$-momentum state.
Minimizing the free energy
\begin{eqnarray}
   &&
   F=E_{\rm g}
    +\sum_k\sum_{\sigma=\pm}\bar{n}_k^\sigma\omega_i^\sigma(k)
   \nonumber\\
   &&\qquad
     +k_{\rm B}T\sum_k\sum_{n^\pm}P_k(n^-,n^+){\rm ln}P_k(n^-,n^+)\,,
   \label{E:F}
\end{eqnarray}
with respect to $P_k$ at each $k$ under the condition (\ref{E:const})
together with the trivial constraints $\sum_{n^\pm}P_k(n^-,n^+)=1$, we
obtain the optimum distribution functions as
\begin{equation}
   \bar{n}_k^\pm
   =\frac{1}
   {{\rm exp}\{[\omega^\pm(k)-\mu(S+s)/\omega_k]/k_{\rm B}T\}-1}
   \,,
\end{equation}
with a Lagrange multiplier $\mu$ due to Eq. (\ref{E:const}), where the
dispersions $\omega^\pm(k)$ are restricted to $\omega_1^\pm(k)$ or set to
$\omega_1^\pm(k)+\omega_0^\pm(k)$ according as we consider linear or
interacting modified spin waves.
In the present formalism, the dispersion relations are determined without
modifying the original Hamiltonian (\ref{E:Hexp}) and then the Lagrange
multiplier is introduced so as to construct a reliable thermodynamics,
which is essentially different from the Takahashi scheme \cite{T168} and
enables us to investigate much wider temperature range.
Otherwise the Schottky peak of the specific heat, for instance, can not be
reproduced at all \cite{Y14008}.
In this context, a mixed Bose-Fermi representation of spin operators
\cite{I1082} may be another useful scheme to thermal calculations.

   The magnetic susceptibility is calculated as
\begin{equation}
   \chi=\frac{(g\mu_{\rm B})^2}{3k_{\rm B}T}
        \sum_{\sigma=\pm}\bar{n}_k^\sigma(\bar{n}_k^\sigma+1)\,,
   \label{E:chi}
\end{equation}
and is shown in Fig. \ref{F:chiT}, together with quantum Monte Carlo
calculations at $N=32$, which are almost the long-chain-limit behavior.
Modified spin-wave calculations well agree to the numerical findings,
covering the low-temperature region to be hardly reached numerically.
The decreasing behavior turns increasing for $k_{\rm B}T\agt 3J$.
A minimum in the susceptibility-temperature product is one of the most
remarkable features of ferrimagnets.
The ferromagnetic and antiferromagnetic terms
$\bar{n}_k^\mp(\bar{n}_k^\mp+1)$ in Eq. (\ref{E:chi}), respectively,
contribute increasing and decreasing behaviors with increasing temperature.
Since the ferromagnetic (antiferromagnetic) features are predominant
for $S>2s$ ($S<2s$) \cite{Y1024}, the present cases with
$(S,s)=(\frac{5}{2},\frac{1}{2})$ are rather biased ferromagnetically.
In order to evaluate the exchange coupling constants in the
Mn(hfac)NIT-$R$ systems, we further compare quantum Monte Carlo
calculations with experimental findings.
The susceptibility measurements on [Mn(hfac)NIT-Ph]$_6$ at
$H=0.5\,\mbox{T}$ \cite{C2795} and those on [Mn(hfac)NIT-$i$-Pr]$_\infty$
under no field \cite{C1756} are also plotted in Fig. \ref{F:chiT}, taking
$J/k_{\rm B}$ to be $370\,\mbox{K}$ and $360\,\mbox{K}$, respectively.
Carrying out semiclassical calculations, Caneschi {\it et al.}
\cite{C1756} estimated $J/k_{\rm B}$ for [Mn(hfac)NIT-$R$]$_\infty$ as
$474.5\,\mbox{K}$ ($R=i\mbox{-Pr}$), $373.3\,\mbox{K}$ ($R=\mbox{Et}$),
$311.8\,\mbox{K}$ ($R=\mbox{Me}$), and $299.5\,\mbox{K}$ ($R=\mbox{Ph}$).
However, smaller values are obtained through a Fischer's model
\cite{F343}.
As for [Mn(hfac)NITPh]$_6$, no quantitative assignment of $J$ has been
given so far.
The present estimates will contribute toward establishing the standard.

\section{Nuclear Spin-Lattice Relaxation}

   NMR measurements on isotropic magnets are necessarily performed with
an applied field.
An induced Zeeman energy gap is usually smaller than the exchange
interaction but larger than the nuclear energy scale:
$10^5\hbar\omega_{\rm N}\alt 10^2g\mu_{\rm B}H\alt J$.
Considering the electronic-nuclear energy-conservation requirement, the
Raman process should play a leading role in the nuclear spin-lattice
relaxation \cite{B359}.
The Raman relaxation rate is generally given by
\begin{eqnarray}
   &&
   \frac{1}{T_1}
    =\frac{4\pi\hbar(g\mu_{\rm B}\gamma_{\rm N})^2}
          {\sum_n{\rm e}^{-E_n/k_{\rm B}T}}
     \sum_{n,m}{\rm e}^{-E_n/k_{\rm B}T}
   \nonumber\\
   &&\ \times
     \big|
      \langle m|\sum_i(A_i^zS_i^z+a_i^zs_i^z)|n\rangle
     \big|^2
     \,\delta(E_m-E_n-\hbar\omega_{\rm N})\,,
\label{E:T1def}
\end{eqnarray}
where
$A_j^z$ and $a_j^z$ are the dipolar coupling constants between
the nuclear and electronic spins in the $j$th unit cell,
$\omega_{\rm N}\equiv\gamma_{\rm N}H$ is the Larmor frequency of the
nuclei with $\gamma_{\rm N}$ being the gyromagnetic ratio, and the
summation $\sum_n$ is taken over all the electronic eigenstates
$|n\rangle$ with energy $E_n$.
Taking account of the significant difference between the electronic
and nuclear energy scales, the relaxation rate (\ref{E:T1def}) is
expressed in terms of modified spin waves as
\begin{eqnarray}
   &&
   \frac{1}{T_1}
   \simeq
   \frac{2\hbar(g\mu_{\rm B}\gamma_{\rm N})^2}{N}
   \sum_{\sigma=\pm}
   \nonumber\\
   &&\ \times
   \Bigl[
    \sum_k|A_0^z\psi^{-\sigma}(k)-a_0^z\psi^\sigma(k)|^2
          \bar{n}_k^\sigma(\bar{n}_k^\sigma+1)\rho^\sigma(k)
   \nonumber\\
   &&\ +
    \sum_k \!'|A_{2k}^z\psi^{-\sigma}(k)-a_{2k}^z\psi^\sigma(k)|^2
          \bar{n}_k^\sigma(\bar{n}_k^\sigma+1)\rho^\sigma(k)
   \Bigr]\,,
   \label{E:T1MSW}
\end{eqnarray}
where $\sum_k \!\!\!\!'$ denotes the limited summation
$\sum_k-\sum_{k=0,\pi}$,
$\psi^\sigma(k)=(S+s)/2[(S-s)^2+4Ss\sin^2(k/2)]^{1/2}+\sigma/2$, and
$A_k^z$ and $a_k^z$ are the Fourier transforms of the coupling constants,
whose $k$ dependences are hereafter assumed to be negligible.
A contrast between zero and one dimensions lies in the spectral density
$\rho^\sigma(k)$, which originates in the energy-conservation requirement
$\delta(E_m-E_n-\hbar\omega_{\rm N})$ in Eq. (\ref{E:T1def}).
In the thermodynamic limit $N\rightarrow\infty$, $\rho^\sigma(k)$ is
definitely the {\it differential coefficients} of the dispersion
relations, while for small clusters, it is approximately replaced by the
{\it difference quotients}:
\begin{mathletters}
   \begin{eqnarray}
   &&
   \rho^\pm(k)=
    \frac{\displaystyle 1}
         {\displaystyle
          \left|
          \left[
           {\rm d}\omega^\pm(k)/{\rm d}k
          \right]_{k=k_0}
          \right|}
   \qquad\ \ \,{\rm for}\ N\rightarrow\infty\,,
   \label{E:rhoN}\\
   &&
   \rho^\pm(k)\simeq
    \frac{\displaystyle 2\pi/N}
         {\displaystyle |\omega^\pm(k)-\omega^\pm(k-2\pi/N)|}
   \ {\rm for}\ N=O(1)\,,
   \label{E:rho6}
   \end{eqnarray}
\end{mathletters}
where $k_0$ as a function of $k$ is given by
$\omega^\pm(k_0)-\omega^\pm(k)-\hbar\omega_{\rm N}=0$.
If we further process Eq. (\ref{E:rhoN}) assuming the predominance of
$k\simeq 0$ contributions \cite{Y2324} in integrating Eq. (\ref{E:T1MSW}),
which is well justified unless temperature and an applied field are
sufficiently high and strong, respectively, we obtain an expression
\begin{equation}
   \rho^\pm(k)
   \simeq\frac{1}{2v\sqrt{k^2+\hbar\omega_{\rm N}/v}}
   \ \ \ {\rm for}\ N\rightarrow\infty\,,
   \label{E:rhoNsimeq}
\end{equation}
where $v=[Ss-(S+s){\mit\Gamma}_1+\sqrt{Ss}{\mit\Gamma}_2]J/2(S-s)$ is the
curvature of the dispersion relations at small momenta.
Equation (\ref{E:rhoNsimeq}) is in contrast with Eq. (\ref{E:rho6}) in
that it depends on an applied field.

   Another consideration should be directed to the modified spin-wave
scheme in calculating the relaxation rate.
For isotropic ferrimagnets with gapless excitations, the constraint
(\ref{E:const}) works so well as not only to suppress the thermal
divergence of the boson number but also to give a precise description of
the low-temperature thermodynamics \cite{Y211}.
On the other hand, once a field is applied and a gap ${\mit\Delta}$ opens
in the electronic energy spectrum, the boson number should exponentially
decreases as $\propto{\rm e}^{-{\mit\Delta}/k_{\rm B}T}$ at low
temperatures, whereas the constraint still keeps it finite even at
$k_{\rm B}T\ll g\mu_{\rm B}H$.
Then we adjust Eq. (\ref{E:const}) to the present situation as
\begin{equation}
   (S+s)\sum_k\sum_{\sigma=\pm}\frac{\bar{n}_k^\sigma}{\omega_k}
  =(S+s)N{\rm e}^{-{\mit\Delta}/k_{\rm B}T}\,.
   \label{E:constH}
\end{equation}
This condition smoothly turns into Eq. (\ref{E:const}) as $H\rightarrow 0$
and the modification is essentially restricted to the sufficiently
low-temperature region $k_{\rm B}T\alt{\mit\Delta}\alt 10^{-2}J$.
\widetext
\vspace*{2mm}
\begin{figure}
\centerline
{\mbox{\psfig{figure=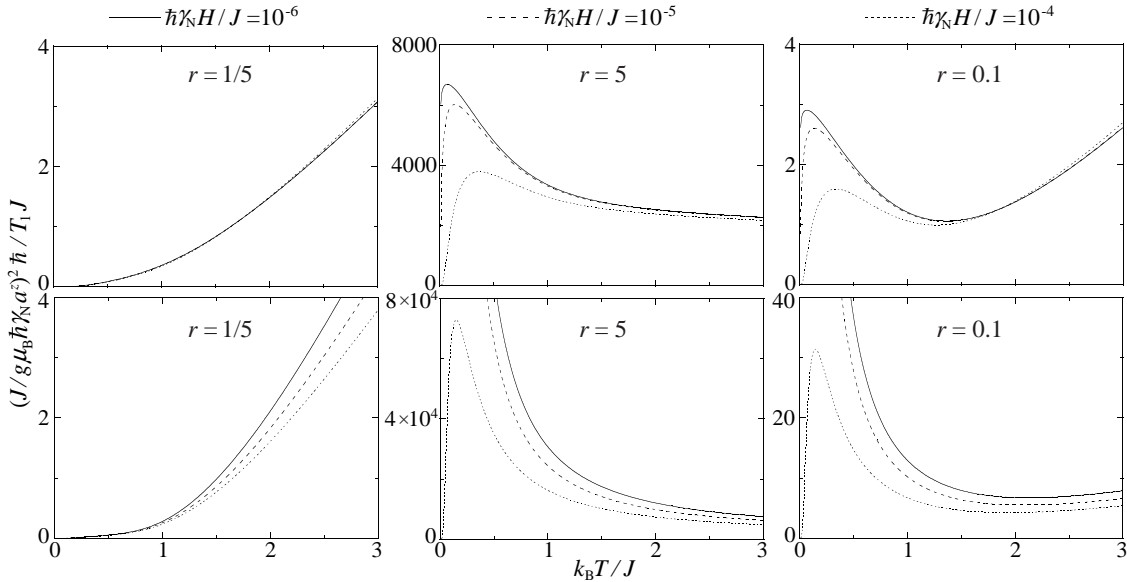,width=150mm,angle=0}}}
\vspace*{2mm}
\caption{The interacting-spin-wave calculations of the temperature
         dependence of the nuclear spin-lattice relaxation rate with
         varying location of the probe nuclei at $N=6$ (the upper three)
         and $N=\infty$ (the lower three).}
\label{F:T1T}
\end{figure}

\begin{figure}
\centerline
{\mbox{\psfig{figure=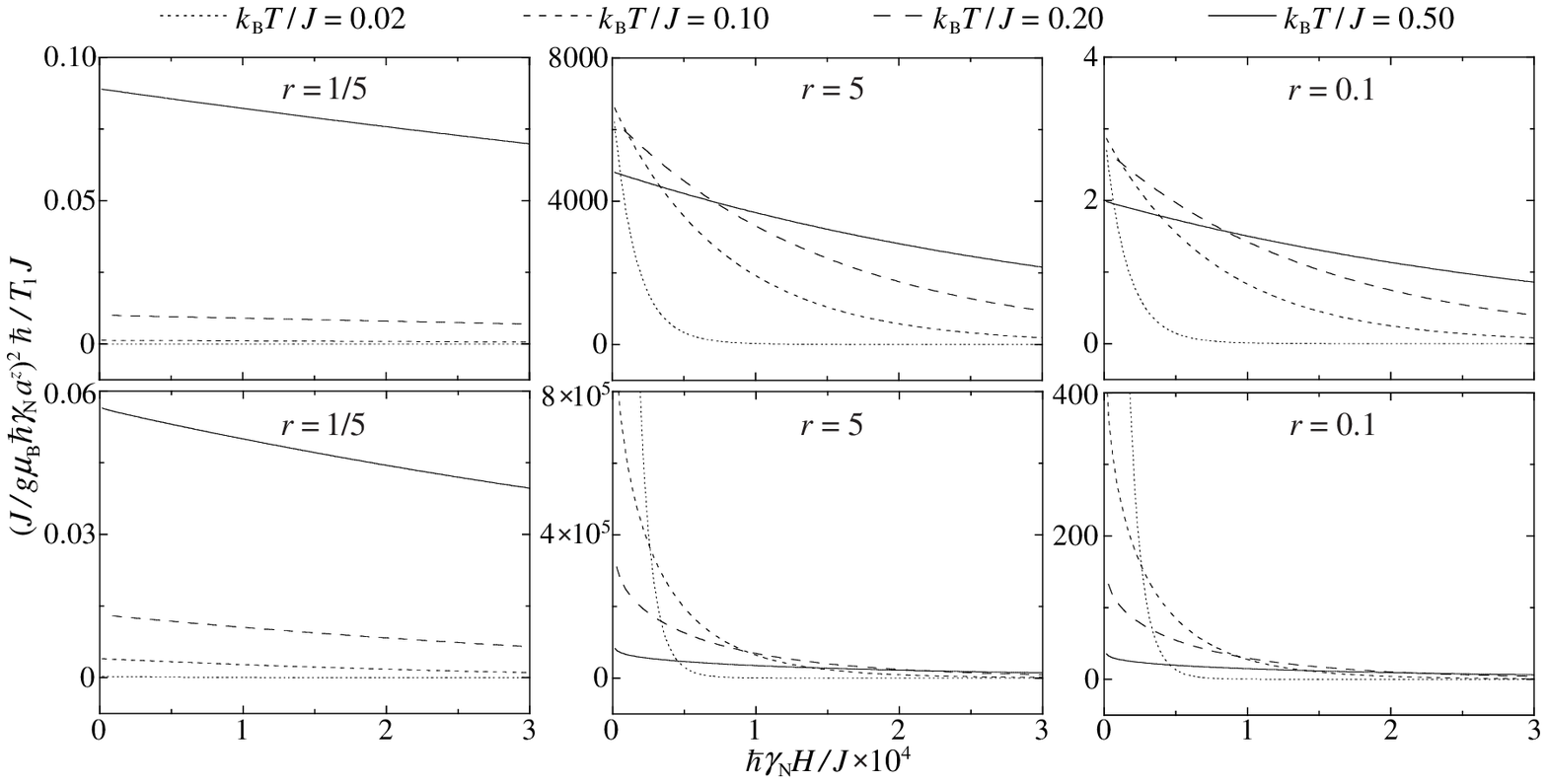,width=156mm,angle=0}}}
\vspace*{2mm}
\caption{The interacting-spin-wave calculations of the field dependence
         of the nuclear spin-lattice relaxation rate with varying location
         of the probe nuclei at $N=6$ (the upper three) and $N=\infty$
         (the lower three).}
\label{F:T1H}
\end{figure}
\vskip 2mm
\narrowtext

\begin{figure}
\centerline
{\mbox{\psfig{figure=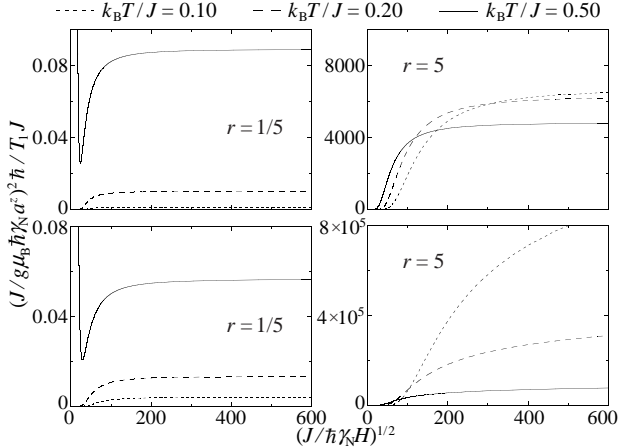,width=82mm,angle=0}}}
\vspace*{1.5mm}
\caption{The interacting-spin-wave calculations of the nuclear
         spin-lattice relaxation rate as a function of inverse square root
         of an applied field with varying location of the probe nuclei at
         $N=6$ (the upper two) and $N=\infty$ (the lower two).}
\label{F:T1Hinv}
\end{figure}

   Now the relaxation rate is calculated for (MnNIT)$_6$ and
(MnNIT)$_\infty$.
Besides temperature and an applied field, it may be a function of the
location of probe nuclei, which can be described by a parameter
$r\equiv A^z/a^z\sim(d_s/d_S)^3$, where $d_S$ and $d_s$ are the distances
between the nuclear and electronic spins.
Figure \ref{F:T1T} shows $1/T_1$ as a function of temperature.
The ferromagnetic and antiferromagnetic spin waves contribute different
temperature dependences to $1/T_1$ and their weights vary with the
parameter $r$.
Considering the predominance of the $k\simeq 0$ contributions in
Eq. (\ref{E:T1MSW}),
{\it at $r\simeq\psi^\sigma(0)/\psi^{-\sigma}(0)=(S/s)^\sigma$, the
$\sigma$ excitation mode hardly mediates the nuclear spin relaxation},
in other words, {\it the electronic excitations of $\sigma$ mode are
invisible to the nuclear spins}.
The temperature dependence of $1/T_1$ is indeed of antiferromagnetic
(ferromagnetic) aspect at $r=1/5$ ($r=5$), while otherwise it is of
mixed aspect.
If the probe nuclei are located as $r\simeq 1/5$, the relaxation rate is
extremely small.
These calculations can be observed by taking different kinds of nuclei as
probes, such as $^1$H, $^{13}$C, $^{19}$F, and $^{55}$Mn in the present
systems.
Equation (\ref{E:T1MSW}) is still valid for $r\rightarrow\infty$, which
corresponds to $^{55}$Mn NMR.

   Figure \ref{F:T1H} shows $1/T_1$ as a function of an applied field.
There appears a clear contrast between zero and one dimensions, where
$1/T_1$ is saturated and diverging, respectively, with decreasing
temperature and field, as long as the ferromagnetic spin waves are
{\it visible} to the nuclear spins.
In finite clusters, any field dependence of $1/T_1$ is necessarily
attributed to $\bar{n}_k^\pm$, which are simply exponential with respect
to $H$, whereas in long chains, $1/T_1$ is more varied with a field,
depending on it through both $\bar{n}_k^\pm$ and $\rho^\pm(k)$.
Therefore, unless the Zeeman energy becomes comparable to the exchange
interaction, $1/T_1$ exhibits little field dependence in finite clusters.
In order to bring out their characteristic field dependences more
quantitatively, we plot $1/T_1$ as a function of $1/\sqrt{H}$ in
Fig. \ref{F:T1Hinv}.
$\bar{n}_k^-$ and $\rho^-(k)$ are both peaked at $k=0$.
More and more weight centers on $k=0$ with decreasing temperature and
field for $N\rightarrow\infty$ in particular.
Therefore, at low temperatures, the $N\rightarrow\infty$ $k$ integration
in Eq. (\ref{E:T1MSW}) may approximately be replaced by the $k=0$
contribution, which is in proportion to $1/\sqrt{H}$.
Thus, as long as an applied field is moderate, {\it a $1/\sqrt{H}$-linear
behavior is observed at low temperatures}.
{\it With increasing temperature it turns logarithmic}, sloping more
gently, due to the $k$-integration effect.
{\it Under strong fields, they are all masked behind the overwhelming
Zeeman term} $\propto{\rm e}^{-g\mu_{\rm B}H/k_{\rm B}T}$ coming from
$\bar{n}_k^-$.
At $r=1/5$, where only the antiferromagnetic spin waves are {\it active}
for $1/T_1$, the nuclear spins exhibit extremely slow dynamics.
Since $\bar{n}_k^+$ is not peaked even at low temperatures, there appears
no $1/\sqrt{H}$ dependence.
With increasing temperature and field, the antiferromagnetic excitation
branch lowers in energy and thermally assists the relaxation, ending up
with increasing $1/T_1$.
{\it A minimum of $1/T_1$ as a function of $H$ can be observed only when
the ferromagnetic spin waves are almost off}.
In finite clusters, the discrete spectrum may in principle lead to $1/T_1$
oscillating as a function of $H$.

\section{Concluding Remarks}

   Motivated by inorganic-organic hybrid compounds, [Mn(hfac)NIT-Ph]$_6$
and [Mn(hfac)NIT-$R$]$_\infty$, we have demonstrated model calculations of
the low-energy spin dynamics in ferrimagnetic clusters and chains.
Temperature dependence of $1/T_1$ drastically varies with the location
of the nuclei in both compounds, though the relaxation time scale is much
larger in zero dimension than in one dimension.
There are special points for the nuclei, characterized as
$(d_s/d_S)^3\sim s/S$, where the nuclear spin relaxation can hardly be
assisted by the low-lying ferromagnetic excitations of the electronic
spins and therefore extremely slow dynamics is observed.

   Field dependence of $1/T_1$ in long chains forms a striking contrast
to that in finite clusters, diverging with decreasing field at low
temperatures.
The present observations should be distinguished from the $1/\sqrt{H}$ or
${\rm ln}(1/H)$ dependence of diffusion-dominated dynamics
\cite{H965,A420}, which originates from transverse spin fluctuations and
distinctly appears at high temperatures.
In the vicinity of the special points of $(d_s/d_S)^3\sim s/S$, a minimum
of $1/T_1$ as a function of $H$ can be observed.

   Besides the Mn(hfac)NIT-$R$ systems,
there are a series of ferrimagnetic bimetallic chain compounds
$M$Cu(pba)(H$_2$O)$_3$$\cdot$$n$H$_2$O
($M=\mbox{Mn},\mbox{Ni}$;
 $\mbox{pba}=1,3\mbox{-propylenebis(oxamato)}$) \cite{P138} and
$M$Cu(pbaOH)(H$_2$O)$_3$$\cdot$$n$H$_2$O
($M=\mbox{Fe},\mbox{Co},\mbox{Ni}$;
 $\mbox{pbaOH}=2\mbox{-hydroxy-1,3-propylenebis(oxamato)}$) \cite{K3325},
which are also describable by the Hamiltonian (\ref{E:H}).
Since their exchange coupling constants are much smaller than those of the
Mn(hfac)NIT-$R$ systems, they are complementary in field-applied
measurements.
We hope that the present calculations will stimulate further experimental
explorations into quantum dynamics on the way from zero- to
one-dimensional magnets.
Nuclear spin-lattice relaxation-time measurements on [Mn(hfac)NIT-Ph]$_6$
and [Mn(hfac)NIT-$R$]$_\infty$ are strongly encouraged.

   This work was supported by the Ministry of Education, Culture, Sports,
Science, and Technology of Japan, and the Nissan Science Foundation.

\widetext
\end{document}